\documentstyle[12pt]{article}
\newcommand{ \be }{\begin{equation}}
\newcommand{ \ee }{\end{equation}}
\newcommand{ \bea }{\begin{eqnarray}}
\newcommand{ \eea }{\end{eqnarray}}

\newcommand{ \la }{\langle}
\newcommand{ \ra }{\rangle}

\newcommand{ \bV }{{\bf V}}

\newcommand{ \bP }{{\bf P}}
\newcommand{ \bp }{{\bf p}}
\newcommand{ \bq }{{\bf q}}

\newcommand{ \ty }{{\tilde{y}}}
\newcommand{ \tV }{{\tilde{V}}}
\newcommand{ \tVp }{{\tilde{V_\perp}}}
\newcommand{ \cQ }{{\cal{Q}}}

\begin{document}
\begin{center}
{\em
Submitted to Phys. Rev. C.
}

\end{center}

\begin{flushright} {
PITT-RHI-96-21\\
\it April 26, 1995
} 
\end{flushright} 

\begin{center}
{\bf
ANISOTROPIC TRANSVERSE FLOW AND THE HBT CORRELATION FUNCTION
}\\
\bigskip
\bigskip
S. A. VOLOSHIN\footnote{ On leave from Moscow Engineering Physics Institute, 
     Moscow, 115409,  Russia}
and W. E. CLELAND

\bigskip
{\em
University of Pittsburgh, Pittsburgh, PA 15260
}\\

\end{center}

\bigskip
{\footnotesize
\centerline{ABSTRACT}
\begin{quotation}
\vspace{-0.10in}
%

Utilizing the Lorentz invariance of the correlation function we study 
the effects of anisotropic transverse flow on the HBT correlation function. 
In particular we show that directed flow would evidence itself by non zero 
``side--long'' and ``side--out'' terms in the correlation function. 
We also show that the study of ratios of the correlation functions evaluated 
on different event subsamples and/or different Lorentz systems provides
information on the source dimensions and velocity which could be less
biased by corrections of different kinds.
We present a fitting technique appropriate for HBT analysis which does
not require binning and is especially useful for multidimensional 
fitting.
\vspace{7pt}
\newline
PACS number(s): 13.85.-t, 13.85.Hd, 25.70.Pq
\end{quotation}}
\bigskip

\newpage
\section{Introduction}

The Hanbury-Brown--Twiss (HBT) analysis of multiparticle production processes
is becoming a widely used technique. It provides  information on the
space-time evolution of an excited strongly interacting system produced
in high energy collisions.
In recent years many experiments have collected high statistics data which
permits detailed multidimensional analysis of the correlation function.
From the theoretical point of view many valuable results have been obtained,
both in the understanding the of the HBT method itself and
in the ways it can be applied 
to the data~\cite{lchap}--\cite{laksi}.
It is now clear, for example, that the correlation function depends
sensitively on the dynamics of the expansion of the system.
Most of the theoretical observations have been made using different kinds 
of models to describe this expansion.

The discovery of  transverse directed flow in the collisions
of ultrarelativistic nuclei~\cite{l877prl} implies that the HBT analysis
of such data should take the effects of flow into account.
In our previous publication~\cite{lvc96} we used RQMD generated events
to show how directed flow affects the extracted HBT radii.
Utilization of the Lorentz invariance of the correlation function, 
defined as the ratio of the invariant two particle distribution
to the product of two invariant one particle distributions,
can provide valuable information about the source and can be used as 
a check of the validity of the assumptions contained in certain models.
These questions are addressed in the current paper mostly in the
context of directed flow.
In particular we argue that directed flow would evidence itself by non zero 
``side--long'' and ``side--out'' cross terms in the correlation function,
even if the correlation function is determined with an azimuthally symmetric
event sample (case in which the reaction plane is not determined).

In dealing with the experimental data we find that the conventional
method of fitting for the correlation function (especially multidimensional
fitting) has several disadvantages and sometimes gives biased results, in spite
of many improvements and suggestions made recently~\cite{lna35}--\cite{lwa80}.
The main problems relate to the low statistics in some bins and to a proper 
evaluation of fluctuations in the distributions of mixed pairs.
Multidimensional fitting is essential for the questions under consideration, 
and we present a fitting method without binning which is free from 
the problems mentioned above and is very flexible in its applications.

The paper is organized in a following way. We start with Lorentz 
properties of the correlation function and discuss how the extracted source
radii depend on the rapidity of the frame in which the analysis is done,
and how anisotropic transverse flow affects the correlation function.
We then discuss the ratio of the correlation functions 
evaluated with different event subsamples and/or in different Lorentz
systems and discuss how these ratios could help in the selection of 
the appropriate functional form for the correlation function
and in the extraction of parameters of the source such as the anisotropic 
flow velocity.
Finally we present the new method of fitting the correlation function 
which does not require binning.

\section{The HBT correlation function in different frames}

The objective of an HBT study of the space-time evolution of multiparticle
production is typically 
the determination of the effective source dimensions,
time duration of particle emission and the velocity of the source.
Boson interferometry is a very effective  tool for this purpose,
but  like any other tool has its own limitations.
The first limitation, a rather trivial one, is that one can study 
only that part of the source which emits particles into the experimental 
acceptance.
This very simple statement mathematically can be written as space-momentum
correlation of the source function. 
It implies that the study of the
correlations between particles produced in different parts of phase space 
permits one to measure the source size as it seen from different directions.

The second limitation is not so trivial. It arises from the physics of
interferometry. 
{\em Interferometry measures the distance between pions at the 
instant the second pion is produced.}
This distance depends not only on the size of the source, but
also the duration of the emission (which affects the distance traversed 
by the first pion before the second pion is produced) and 
the velocity of the source (the source could move to another place 
before the production of the second pion).
Mathematically it is related to the fact that only three of the four
components of the 
pair momentum difference $q_i$ are independent, because of the particles
being on the mass shell.
The fourth component of the momentum difference
can be expressed through first three components and the pair velocity $\bV$:
\be
q_0=\bV \bq.
\label{emass}
\ee
Often this results in the situation when there are more unknowns than the number
of parameters possible to extract experimentally, and additional assumptions
must be made about the source to resolve the problem.

The correlation function, being a ratio of invariant distributions, 
is invariant with respect to a Lorentz transformation:
\be
C(\bq,\bP)=\frac{d^6n/d^3p_1 d^3p_2}{d^3n/d^3p_1 d^3n/d^3p_2}
=C(\bq',\bP'),
\label{ecorrfun}
\ee
where $\bP=\bp_1+\bp_2$ is the total momentum of the pion pair, and
the prime denotes
the values in any other Lorentz system.
Lorentz invariance provides us with the possibility to calculate 
the correlation function in any frame provided we know its functional form
and value in any other frame. 
Below we use the following notation: by an asterisk ($^*$) we denote
the values in the source rest frame (but we use $y^*$ for
the source rapidity in the laboratory frame), and a tilde ($\,\tilde{}\,$)
is used for the LCMS (Longitudinally Co-Moving System) frame,
in which the longitudinal velocity of the pair is equal to
zero ($\tV_z=0$).

For example, using this notation, the Lorentz transformations 
between the analysis frame (with rapidity $y$)
 and the source rest frames reads:
\be
q^*_z=q_z \cosh(y-y^*)+q_0 \sinh(y-y*),
\ee
\be
q^*_0=q_0 \cosh(y-y^*)+q_z \sinh(y-y*).
\ee
The longitudinal and transverse velocity components of the pair in 
the analysis frame are:
\be
V_z=\tanh(\ty-y),
\ee
\be
V_{\perp}=\tilde{V_\perp}/\cosh(\ty-y).
\ee

Very often the correlation function is written in the form:
\be
C(q)=1+\lambda e^{-\cQ(q)}.
\ee
Using the  small $q$ approximation in the source rest frame (we discuss below 
the existence of such a frame ) one can represent $\cQ$ as:
\be
\cQ=a_0 q^{*2}_0 +a_x q^{*2}_x +a_y q^{*2}_y +a_z q^{*2}_z.
\label{egauss1}
\ee
Our notation differs here from that often found in the literature
($a_0=\tau^2;\;a_x=R_x^2;\;a_y=R_y^2;\;a_z=R_z^2$). 
We have introduced the new notation 
since the fits to experimental data can give negative values for
these parameters, especially if there are cross-terms.
Throughout this paper we use the coordinate system in which 
the $z$ axis is along the beam direction,
the $x$ axis is along the ``out''~\cite{lbertsch} direction 
(the direction in the transverse plane along the momentum of the pair) 
and the $y$ axis is along the ``side'' direction, which is 
perpendicular to the ``out'' and beam directions.
Note that in the source rest frame the corresponding cross terms
$a_{x0}= a_{y0}= a_{z0}=0$, if
the source velocity is defined as~\cite{lvc96}:
\be
v_i =\frac{\la r_i t \ra - \la r_i \ra \la t \ra}
        {\la t^2 \ra - \la t \ra ^2 }                  =0.
\ee
We also {\em assume} that the source is azimuthally symmetric (in the source rest
frame): $ a_{xy}=a_{yz}=0,\,a_x=a_y=a_\perp$, and there is no correlation
of the ``$x$-$z$'' type: $a_{xz}=0$.

As was mentioned above only three components of $q_i$ are independent,
thus the experimental data can be fitted by the correlation function
in the form:
\be
\cQ=A_x q_x^{2} + A_y q_y^{2} + A_z q_z^{2} + A_{xz} q_x q_z.
\label{egauss2}
\ee
The correspondence of our parameters $A$ to the often used
$R^2$ is obvious, and we have adopted again a new notation 
for the reason given above.
We assume that the source can move only longitudinally
(we consider the more general case in the following section)
and make use of the Lorentz invariance of the correlation function.
After some algebraic manipulations one finds
the relations between the parameters $A$, which are to be extracted from
the data analyzed in the analysis frame defined by rapidity $y$, 
and the source parameters $a_i$ introduced above:
\bea
A_z &=& \frac{1}{\cosh^2(\ty-y)}[a_z \cosh^2(\ty-y^*) +a_0 \sinh^2(\ty-y^*)]
\label{eab}
\\
A_x &=& a_\perp + \frac{\tilde{V_\perp}^2}{\cosh^2(\ty-y)}
         [a_0 \cosh^2(y-y^*) +a_z \sinh^2(y-y^*)]
\\
A_y &=& a_\perp
\\
A_{xz} &=& \frac{2 \tilde{V_\perp}^2}{\cosh^2(\ty -y)}
            [a_z \cosh(\ty-y^*)\sinh(y-y^*)
\nonumber
\\
& &
+ a_0 \cosh(y-y^*)\sinh(\ty-y^*)].
\label{eae}
\eea
In the case when the analysis frame coincides with
 the source rest frame ($y=y^*$) these equations 
(\ref{eab}--\ref{eae}) give the well known equalities:
\bea
A_z &=& a_z+a_0 V_z^2,
\\
A_x &=& a_\perp +a_0 \tVp^2,
\\
A_y &=& a_y,
\\
A_{xz}&=& 2a_0 \tVp V_z.
\eea

When the analysis frame coincides with the LCMS ($y=\ty$) we have:
\bea
A_z &=& a_z \cosh^2(\ty-y^*)+a_0 \sinh^2(\ty-y^*),
\\
A_{xz}&=& 2 \tVp \sinh(\ty-y^*)[a_z-a_0],
\\
A_x-A_y&=& \tVp^2[a_0\cosh^2(\ty-y^*)+a_z \sinh^2(\ty-y^*)],
\eea
and it follows:
\bea
A_z-\frac{A_x-A_y}{\tVp^2} &=& a_z -a_0,
\\
\frac{A_{xz}/\tVp}{A_z+(A_x-A_y)/\tVp^2} &=& \tanh(2(\ty-y)).
\eea
The last equation, in principle, permits to evaluate 
the velocity of the source with respect to the pair velocity.

The equations presented above can be used as a check of the validity of
the functional form (Gaussian) used for the correlation function.
One needs to evaluate $A_i$ as a function of
analysis frame rapidity $y$ 
and compare the observed dependence of the parameters
 on the rapidity with the dependence given by these expressions.
One of the simplest tests of this type would
be to look for the independence of the parameter $A_y$
(often called  the ``side radius squared'')
on the analysis frame rapidity $y$.

In the discussion above it is assumed that the source rest frame exists, 
that is that all pions entering the detector are emitted by a source
with fixed rapidity. In general this is not true. 
It is rather likely that the  effective source
should be described as a superposition of sources moving with 
different rapidities. In this case, approximating 
the correlation function by a single (multidimensional) Gaussian
will not be successful; it could result, for example, in the dependence 
of $A_y$, the effective source ``side'' size, (and/or the coherence
parameter $\lambda$) on the analysis frame rapidity.
This arises from the fact that the radii $A_i$ (other than $A_y$)
depend on the analysis frame rapidity $y$ (see Eq.\ref{eae}).
If the source parameter $a_y$ is different 
for the different sources, a fit to a single Gaussian 
results in an apparent $y$-dependence for the $A_y$.
Note that in the saddle point approximation~\cite{laksi}
the effective pion source is exactly in the (Gaussian) form considered above. 
In this case the question about the existence of the source rest frame
is equivalent to a question about the validity 
of the saddle point approximation.

\section{Directed flow effects}

Here we discuss how directed flow affects the terms in 
the correlation function related to the ``side'' direction.
In particular we argue that in the presence of directed flow the ``side--long''
and ``side--out'' cross terms of the correlation function become non zero.
The experimental measurements of these terms (especially as a function
of centrality of the collision) would be of great interest.

We start with the simple case when the correlation function can be written in
the ``source rest frame'' in a pure Gaussian 
form (\ref{egauss1}--\ref{egauss2})
without any cross terms except the trivial ones related to the fact
that particles are on the mass shell.
We assume also that directed flow exhibits itself simply as a movement
of the source in the transverse plane.
Under this assumption one can calculate the correlation
function in the analysis frame using Lorentz transformations.
To do this we start with the expression for the correlation function
in the source rest frame(\ref{egauss1}--\ref{egauss2}), 
then perform a Lorentz shift in the
direction of the transverse flow (with transverse rapidity $y_\perp$),
and then do a Lorentz shift along the $z$ axis to the frame moving 
with rapidity $y$. 
In the final step  we use the mass shell constraint, Eq. (\ref{emass}). 
We denote by $\psi$ the angle between the direction of the
source velocity (the reaction plane)
and the $x$ axis (``out'' direction).

This gives the expressions:

\begin{eqnarray}
A_z&=&\frac{1}{\cosh^2(\ty-y)}(a_z \cosh^2(\ty-y^*) 
                +a_0\sinh^2(\ty-y^*)\cosh^2 y_\perp
\nonumber \\
& &\,\,\,\;\;\;\;\;   +a_\perp \sinh^2(\ty-y^*)\sinh^2 y_\perp),
\label{eaab}
\\
A_x &=& a_\perp \sin^2\psi +a_\perp (\cos\psi \cosh y_\perp
+\frac{\tVp \cosh(y-y^*)\sinh y_\perp}{\cosh(\ty-y)})^2+
\nonumber \\
& & + a_0(\frac{\tVp \cosh(y-y^*)\cosh y_\perp}{\cosh(\ty-y)}
+\cos\psi \sinh y_\perp)^2 
+a_z \frac{\tVp^2 \sinh^2(y-y^*)}{\cosh^2(\ty-y)},
\\
A_y&=& a_\perp(\cos^2\psi +\sin^2\psi \cosh^2 y_\perp)+
        a_0\sin^2\psi \sinh^2 y_\perp ,
\\
A_{xy}&=& 2 a_\perp \sin\psi \cosh y_\perp
(\cos\psi  \cosh y_\perp + \frac{ \tVp  \cosh(y-y^*) \sinh y_\perp}
{ \cosh(\ty-y)})
\nonumber \\
& &+2a_0 \sin\psi  \sinh y_\perp  ( \frac{\tVp  \cosh(y-y^*) \cosh y_\perp }
{ \cosh(\ty-y)}+ \cos\psi \sinh y_\perp )
\nonumber \\
& & - a_\perp \sin (2\psi),
\\
A_{xz}&=& 2 a_\perp  \sinh y_\perp \frac{ \sinh(\ty-y^*)}{ \cosh(\ty-y)}
(  \cos\psi \cosh y_\perp +\frac{ \tVp  \cosh(y-y^*)  \sinh y_\perp }
{ \cosh(\ty-y)})
\nonumber \\
& & + 2a_0   \cosh y_\perp \frac{ \sinh(\ty-y^*)}{ \cosh(\ty-y)}
( \cos\psi \sinh y_\perp +\frac { \tVp  \cosh(y-y^*)  \cosh y_\perp }
{ \cosh(\ty-y)})
\nonumber \\
& &+2a_z \tVp   \sinh(y-y^*)\frac{ \cosh(\ty-y^*)}{\cosh^2(\ty-y)},
\\
A_{yz}&=& (a_\perp +a_0)\frac{\sinh(2y_\perp) \sin\psi  \sinh(\ty-y^*)}
{ \sinh(\ty-y) }.
\label{eaae}
\eea  

Let us consider, for example, the ``side--long'' cross term $A_{yz}$.
Averaging it directly over the reaction plane angle $\psi$ gives zero,
 but the expression 
one has to average is not $A_{yz}$, but the correlation function  
$\exp(-A_{yz}q_yq_z)$. Then directed flow
effectively results in nonzero cross terms in the correlation function.
Note that analogous terms appear also in $A_x,\, A_{xy},\,A_{xz}$.

If no selection of events is done with respect 
to the reaction plane orientation, averaging over $\psi$ gives:
\be
\frac{1}{2\pi}\int d\psi \exp(-A_{yz}q_y q_z)= I_0(B_{yz}q_yq_z),
\ee
where
\be
B_{yz}= (a_\perp +a_0)\frac{\sinh(2y_\perp)  \sinh(\ty-y^*)}
{ \sinh(\ty-y) },
\ee
and one could fit the data using the corresponding form 
of the correlation function.

We should mention that strictly speaking the averaging over the flow angle
should be performed separately in the numerator and denominator 
in the definition of the correlation function, Eq.~(\ref{ecorrfun}), 
rather than averaging the correlation function itself.
The above result is valid if the width of the two particle correlation
 related just to the movement of the source
(this correlation has nothing to do with the quantum interference
of the identical particles) is much wider than the width 
of the correlation due to  quantum interference.
This problem/assumption is relevant not only to anisotropic flow effects.
One should not forget that the HBT correlations
are always ``on top'' of any long and/or short range correlations due to
resonance decays, source movement, energy and momentum conservation, etc..
By fitting the correlation function in a form appropriate only for
quantum interference correlation we implicitly assume
that all other correlations have much greater widths.

Formulae~(\ref{eaab}--\ref{eaae}) can also be useful 
for purposes not directly related to transverse flow.
For example, the case of $\psi=0$ and 
$\tanh(y_\perp)=V_\perp \cosh(\ty-y^*)$ would
give the parameters of the correlation function as measured in the pion
pair rest system.

\section{The ratio of the correlation functions}

We now suggest a method to study the functional form of the correlation
function, exploiting its Lorentz transformation invariance.
First we hypothesize a particular functional form of the correlation function.
We then calculate the parameters in different Lorentz systems and
compare the observed dependence on the rapidity of the analysis frame
with the expected dependence.
In other words we choose the form and parameters of the correlation function
and test for its Lorentz invariance.

In practice we suggest the fitting of the ratio of
the correlation functions:
\be
R(a_x,a_y,a_z,a_0;\lambda,v_x,v_z) 
 \equiv \frac{dN^{(1)}_{true}}{dN^{(2)}_{true}},
\ee
where (1) and (2) denote different event subsamples and/or different Lorentz
systems.
Note that in this method there is no need to produce mixed pairs;
it also much less affected by the corrections of various types (such as
the Coulomb correction).

To illustrate the method
we consider an example related to directed flow.
One can assume that the source of pions in events with different reaction
plane orientation differs only in orientation of directed flow velocity,
but all other parameters describing the source are the same.
Then one can perform a fit to extract the source transverse velocity.
One should fit the ratio:
\be
R(a_x,...)=\frac{1+\lambda e^{-\cQ'}}{1+\lambda e^{-\cQ}},
\ee
where $\cQ'$ and $\cQ$ differ only by a Lorentz shift in the flow direction.
For example, if one can select two event subsamples with 
the flow direction pointing along the
 $+x$ direction and the $-x$ direction, 
then   $\cQ'$ and $\cQ$
would be defined by formulae (\ref{eaab}--\ref{eaae}) with
$\cos\psi$ equal to $+1$ and $-1$, respectively.

One could argue that the source might look different from the $+x$
and  $-x$ direction~\cite{lvc96}, that is the parameters of
the effective sources emitting pions in these directions are different.
In this case it could be better to
use event subsamples corresponding to flow directed along  the $+y$
and  $-y$ directions. In this case, because of symmetry, the only difference
between  $\cQ'$ and $\cQ$ is due to the direction of the movement of
the source.

Note that a simple test of whether the parameters of the source
are the same for different event subsamples (selected, for example, according
to the orientation of the pair momentum with respect to the 
reaction plane) could be the evaluation of 
the correlation function using invariant variables. The simplest
case would be a one dimensional fit in $Q_{inv}$.
If the parameters of the source for different event subsamples 
are the same, and the only difference between them is the
direction of the motion of the source,
the extracted $R_{inv}$ should be the same
for all event subsamples, due to the Lorentz invariance of the correlation
function.
\section{Likelihood function fitting}

The usual approach for the fit of the correlation function
assumes the introduction of binning in some part or even over
the entire phase space.
This approach has a number of disadvantages. It is sensitive to 
the experimental acceptance and to  cuts introduced during the analysis. 
Often some bins have a low number of hits,
which introduces biases in the results. 
Sometimes the parameterized correlation function has 
only a few parameters, but the complicated functional form of the
parameterization may demand binning in many dimensions.

It is attractive in this case to use 
the maximum likelihood fitting technique,
which is free from almost all of the problems mentioned above.
However a direct application of this technique is difficult, because 
in this approach one needs to parameterize not only the correlation
function itself (the goal of the fit), but of the distributions
of ``true'' and ``mixed'' pairs.
The latter is extremely difficult, since one must take into account
numerous nontrivial acceptance and analysis cuts.

We  propose a method based on the likelihood function,
which allows one to avoid these problems.
For simplicity we consider one dimensional case, but the 
generalization of the method for multiple dimensions is straightforward.
In the one dimensional case the correlation function is written as:
\be
C(x) \propto (\frac {dN_{true}}{dx})
            /
    (\frac{dN_{mixed}}{dx}).
\ee
Let us denote the probability density for mixed pairs as
$p(x)$:
\be
\frac{1}{N_{mixed}} \frac{dN_{mixed}}{dx} = p(x)
\ee
Then the  probability density for the true pair distribution can be written as
\be
\frac{1}{N_{true}} \frac {dN_{true}}{dx} = G[C] C(x)p(x),
\ee
where  $G$ is the normalization factor, which is a functional of $C(x)$,
and therefore depends on the parameters which determine $C$.

The log-likelihood function depends on the  
correlation function in the following way:
\be
\ln L[C]=\sum_{k=\{true\ pairs\}} \{ \ln G [C]+\ln C(x_k)+\ln p(x_k)\}.
\ee
The last term in the sum does not depend on $C(x)$ and can be omitted.
But one must evaluate the normalization factor $G[C]$
for different sets of  parameters. 
The problem is that for this calculation one needs to
know the probability density $p(x)$.
The trick of our  method is that the required integral can be evaluated
using ``mixed'' pair distribution.
\bea
G^{-1}[C] 
&=& \displaystyle 
\int dx C(x)p(x) = \int dx C(x) \frac{\textstyle 1}{\textstyle N_{mixed}}
                     \frac{\textstyle dN_{mixed}}{\textstyle dx}
\\
&\approx& \frac{\textstyle 1}{\textstyle N_{mixed}} 
\displaystyle \sum_{i=\{mixed\ pairs\}} C(x_i)
\eea
For example, if we consider the Gaussian form of the correlation function
\be
C(x) = 1+\lambda \exp(-a x^2),
\ee
the log-likelihood function can be written as
\bea
\ln L(a,\lambda)&=&\sum_{k=\{true\  pairs\}} \ln (1+\lambda \exp (-a x_k^2))
\\
& &-N_{true} \ln \{
\frac{1}{N_{mixed}} \sum_{i=\{mixed\  pairs\}} (1+\lambda \exp (-a x_i^2))\}.
\eea

The method has many advantages. First, it gives a  correct evaluation of the
errors in the fit, because it automatically takes into account the fact that
the pairs in the mixed event technique are not independent and in general 
the fluctuations are proportional to
$\delta N_{mixed} \propto N_{mixed}^{3/4}$~\cite{lmixed}, 
in contrast  to the fluctuations in true pairs distributions 
$\delta N_{true} \propto N_{true}^{1/2}$. 
Taking into account that the maximum number of mixed pairs which
can be generated is proportional to 
 the square of the number of events 
($N_{mixed} \propto N^2_{events} \propto N_{true}^2 $),
it follows that the fluctuations in mixed pairs distributions {\em never}
become much less than that in the real pairs distribution, regardless of how
many of them are generated.
Other advantages of the likelihood technique is that it is free from
the problems associated with  finite bin size, and it is very flexible in its 
applications (For example, it is easy to introduce
analysis or acceptance cuts,  and it is very straightforward to use
 different functional
forms of the correlation function.)
\section{Conclusion}

We have presented new techniques for the analysis of like-particle
pairs using the HBT method and for testing the form of the correlation
function used in such studies. Starting from the Lorentz-invariant property
of the correlation function, we derive relationship for the fitted parameters
as a function of the rapidity of the analysis frame. This technique is
particularly useful, but not limited to, studies of interacting systems
in which anisotropic transverse flow is present.

Our results may be summarized as follows:
(i) The analysis of the same data in different Lorentz systems 
helps in the extraction of the source parameters and assessing the
validity of the approximations used, for example the validity of Gaussian
source (saddle point approximation).
(ii) Anisotropic transverse flow would result in the non zero 
``side'' cross terms in the correlation function. 
(iii) The study of ratios of the correlation functions
evaluated in different Lorentz frames (the ratios of ``true''
pair distributions) is an important new analysis tool.
(iv) We have presented a new fitting technique based on the likelihood
function which does not require
binning and has many advantages in comparison to the conventional method.

\section*{Acknowledgments}

The authors acknowledge fruitful discussion with our colleagues from
the E877 Collaboration. The financial support provided by US DOE
under Contract No.~DEFG02-87ER40363.

\end{document}